\documentclass[prb,twocolumn,showpacs]{revtex4}

\usepackage{amsmath,amsfonts,amssymb,bm}

\renewcommand*{\r}[0]{{\bm r}}
\newcommand*{\s}[0]{{\bm s}}
\renewcommand*{\k}[0]{{\bm k}}
\newcommand*{\q}[0]{{\bm q}}
\newcommand*{\p}[0]{{\bm p}}

\renewcommand*{\c}[1]{c^{\vphantom{\dagger}}_{#1}}
\renewcommand*{\d}[1]{d^{\vphantom{\dagger}}_{#1}}
\newcommand*{\cd}[1]{c^\dagger_{#1}}
\newcommand*{\dd}[1]{d^\dagger_{#1}}

\newcommand*{\e}[2]{\epsilon_{#1}^{\rm #2}}
\newcommand*{\m}[2]{{\rm M}_{#1}^{\rm #2}}
\renewcommand*{\v}[4]{{\rm V}_{#1,#2}^{\rm #3,#4}}
\newcommand*{\w}[4]{{\rm W}_{#1,#2}^{\rm #3,#4}}
\newcommand*{\D}[4]{{\cal D}_{#1,#2}^{\rm #3,#4}}

\begin{document}
\bibliographystyle{apsrev}

\title{Ab-initio calculation of optical absorption in semiconductors:\\
       A density-matrix description}

\author{Ulrich Hohenester} 
\email{ulrich.hohenester@uni-graz.at}

\affiliation{Institut f\"ur Theoretische Physik,
  Karl--Franzenz--Universit\"at Graz, Universit\"atsplatz 5, 
  8010 Graz, Austria
  }

\date{May 31, 2001}

\begin{abstract}

We show how to describe Coulomb renormalization effects and dielectric
screening in semiconductors and semiconductor nanostructures within a
first-principles density-matrix description. Those dynamic variables
and approximation schemes which are required for a proper description
of dielectric screening are identified. It is shown that within the
random-phase approximation the direct Coulomb interactions become
screened, with static screening being a good approximation, whereas the
electron-hole exchange interactions remain unscreened. Differences and
similarities of our results with those obtained from a corresponding GW
approximation and Bethe-Salpeter equation Green's function analysis are
discussed.

\end{abstract}



\maketitle


\section{Introduction}

The study of the optical properties of nanoscale semiconductor
structures, such as small clusters, nanocrystals or polymer chains, is
acquiring increasing importance also in view of optoelectronic
applications. In these systems, however, the theoretical description of
the optical response faces additional difficulties with respect to
extended crystals, because correlation effects are known to be
enhanced. Besides the single-particle energies and wave-functions for
electrons and holes, also an appropriate description of Coulomb
interactions between electrons and holes is required. For this reason
semiempirical or approximate methods (e.g., based on the effective-mass
approximation) were often the only feasible treatment so far. In many
cases of interest, however, such methods are not sufficiently accurate
and predictive.

A strong impulse towards {\em ab-initio}\/ approaches is recently
coming from the success of standard Density Functional Theory (DFT)
methods to treat extended three-dimensional systems,~\cite{dreizler:90}
with unit cells that may now include thousands of atoms. In
semiconductors, however, such DFT approaches are known to account only
insufficiently for Coulomb correlation effects. The main reason for
this shortcoming is that the dynamics of electron-hole excitations is
not only governed by the DFT single-particle states but also by
polarization effects, which are not fully accounted for in a standard
DFT calculation. More specifically, in semiconductor carriers polarize
their surrounding; in turn, this induced polarization affects both the
propagation of carriers (quasiparticle renormalizations) and the
Coulomb attraction between optically excited electrons and holes
(dielectric screening).

Within the framework of Green's functions it has been known for a long
time how to correct for such polarization screening. For instance,
Hedin and coworker (see Refs.~\onlinecite{hedin:69,aryasetiawan:98} for
a review) showed how to consistently develop a set of equations
accounting for Coulomb correlation effects. In lowest order, this set
of equations reduces to the calculation of the screened Coulomb
potential ${\rm W}$, obtained from the solution of Dyson's equation for
${\rm W}$ within the random-phase approximation, and the self-energy
$\Sigma$ which is calculated within the so-called GW approximation
(GWA).~\cite{hedin:69,aryasetiawan:98} While the GWA approach provides
a strongly improved description of single electrons and holes, the
description of optical absorption requires in addition a proper
treatment of the attractive Coulomb interaction between the optically
excited electron-hole pair. This electron-hole correlation is
conveniently accounted for by solving the Bethe-Salpeter equation,
which describes the propagation the two-particle Green's function (for
details see Ref.~\onlinecite{rohlfing:00} and references therein).

On the other hand, within the last couple of years phenomenological
semiconductor optics, i.e., the simplified bandstructure and dielectric
screening-description of semiconductors,~\cite{haug:93,axt:98} has
become an extremely vivid and successful field of research. Here, the
main interest is on the description of the nonlinear optical response
of semiconductors and semiconductor nanostructures, and of optically
induced coherence effects. A particularly transparent description
scheme for such systems is provided by the dynamics-controlled
truncation of the density-matrix hierarchy:~\cite{axt:98} Contrary to
the Green's function approach, which describes by means of many-body
perturbation theory the {\em dynamic}\/ propagation of quasiparticles,
within the framework of density matrices one selects an appropriately
chosen set of {\em one-time}\/ correlation functions. If Coulomb
interactions conserve the number of electron-hole pairs, it becomes
possible to rigorously classify these correlation functions in powers of
the electric field of the exciting laser,~\cite{axt:98} thus providing
a consistent description scheme for the optical properties of
semiconductors.

If, instead of adopting from the beginning a simplified semiconductor
description, we start from the results of a first-principles
calculation, the most interesting question arises: How is it possible
to describe dielectric screening and Coulomb correlation effects within
a first-principles density-matrix description? To the best of our
knowledge, up to now this problem has not been addressed in the
literature. Its answer, however, would be very interesting for a number
of reasons. First, it would be helpful to establish a link between the
fields of semiconductor optics and of first-principles calculations;
more specifically, such synthesis would allow the first-principles
calculation of non-linear optical properties of semiconductors. Second,
the framework of density matrices requires only some basic knowledge of
quantum mechanics and statistical physics---in contrast to the
framework of Green's functions, which rests on a large machinery of
definitions and auxiliary functions. Thus, one might hope that certain
types of approximations become more transparent; furthermore, the
respective strengths of the two approaches might be helpful in
establishing approximation schemes beyond the presently used
approaches.

This paper is devoted to the discussion of a first-principles
calculation of dielectric screening within the framework of density
matrices. It will be shown that non-pair conserving Coulomb couplings,
whose neglect is at the heart of the dynamics-controlled truncation
approach, have to be retained for a proper first-principles
density-matrix description of dielectric screening and quasiparticle
renormalizations in optically excited semiconductors. Adopting the
usual random-phase approximation and keeping our analysis at the lowest
level of approximation, we will propose a method for the {\em
ab-initio}\/ calculation of optical absorption which closely resembles
that of a corresponding Green's function analysis: first, the exciton
energies $E_x$ and wavefunctions $\Psi^x$ are obtained from the
solutions of the two-particle electron-hole Schr\"odinger equation,
where the electron and hole interact with each other through the
screened direct Coulomb term and the unscreened exchange one; second,
the optical transition energies are given by $E_x$ whose oscillator
strengths are calculated from $\Psi^x$. As the main difference with
respect to the corresponding Green's function analysis, our
prescription suggests to use static screening in the calculation of the
direct Coulomb term.

We have organized this paper as follows. In Sec. II we present our
model Hamiltonian and introduce the framework of density matrices; a
short overview over one of the most widely used truncation schemes in
phenomenological semiconductor optics, the dynamics-controlled
truncation, will be given. Sec. III is devoted to the analysis of
dielectric screening within the framework of density matrices; the
pertinent dynamic variables and approximation schemes are identified.
To keep our analysis as simple as possible, details of our calculations
are postponed to Appendices A and B.  Finally, in Sec. IV we compare
our results with those obtained within the GWA and
Bethe-Salpeter-equation approach and draw some conclusions.  FInally,
in Sec. V we draw some conclusions and summarize our prescription for
the {\em ab-initio}\/ calculation of optical absorption in
semiconductors.


\section{Density matrices}

\subsection{Hamiltonian}\label{sec:ham}

Let us assume that the single-particle energies $\epsilon_a$ and
wavefunctions $\phi_a(\r)$ of the semiconductor system under
investigation are obtained from a DFT calculation. Although we shall
not be too specific about the details of the underlying {\em
ab-initio}\/ approach, we assume that the Hartree and exchange
interactions are included without approximations;~\cite{stadele:97} a
discussion of differences with respect to the local-density
approximation (LDA) will be given below. Let $H_o$ denote that part of
the Hamiltonian which includes the kinetic energy and the lattice
interactions, and $\bar H'$ that part of the electron-electron Coulomb
interactions $H'$ which are included through the Hartree, exchange, and
correlation potentials $v_H+v_x+v_c$. Then, $\epsilon_a$ and
$\phi_a(\r)$ are the DFT eigenstates of $\bar H_o=H_o+\bar H'$, and
$\delta H=H'-\bar H'$ are those Coulomb interactions which are not
included in the DFT approach (and which will be treated in this paper
by means of perturbation theory). The basic Hamiltonian describing the
carrier system is thus of the form:

\begin{equation}\label{eq:ham}
  H=\bar H_o+H_{\rm op}+\delta H,
\end{equation}

\noindent with the single-particle Hamiltonian $\bar H_o$; the
light-semiconductor coupling $H_{\rm op}$; and the Coulomb interactions
$\delta H$ which are not included in DFT. Note that for simplicity in
Eq.~(\ref{eq:ham}) we have neglected all types of carrier-phonon
interactions or other environment couplings. In the spirit of the GWA
approach, in this work we consider in the subtractions to $\delta H$
only the Hartree and exchange terms, thus ignoring possible double
counting of Coulomb correlations $v_c$. Hence, with the Fermionic
field operators ${\bm\psi}(\r)$, which create an electron at position
$\r$, we obtain:

\begin{eqnarray}\label{eq:ham.coul(psi)}
  \delta H&\cong&\mbox{$\frac 1 2$}
  \int d(\r\r')\;\frac{{\bm\psi}^\dagger(\r){\bm\psi}^\dagger(\r')
                       {\bm\psi}(\r'){\bm\psi}(\r)}{|\r-\r'|}\nonumber\\
  &-&\int d(\r\r')\;[v_H(\r)\delta(\r-\r')+v_x(\r,\r')]
  {\bm\psi}^\dagger(\r'){\bm\psi}(\r).\nonumber\\
\end{eqnarray}

In the DFT semiconductor groundstate all states below the fundamental
bandgap are occupied (valence-band states) and all states above the
bandgap are unoccupied (conduction-band states). Next, we transform to
the electron-hole picture and introduce the Fermionic field operators
$\cd{}$ and $\dd{}$, where: $\cd{1}$ creates an electron in the
conduction-band state $\phi_1^{\rm e}(\r)$ with energy $\e{1}{e}$
(henceforth we use numbers $1,2,\dots$ to label single-particle
states); $\dd{2}$ creates a hole in the valence-band state $\phi_2^{\rm
h}(\r)$ with energy $\e{2}{h}$. Since the hole describes the properties
of a missing electron in the valence bands, we adopt the usual
definitions that for a valence band state $v$ the corresponding hole
state $2$ is related through $\e{2}{h}=-\epsilon_v$ and $\phi_2^{\rm
h}(\r)=[\phi_v(\r)]^*$ (with a properly chosen zero-point energy).
With these operators: the single-particle Hamiltonian is of the form
$\bar H_o= \sum_1\e{1}{e}\; \cd{1}\c{1}+\sum_2\e{2}{h}\; \dd{2}\d{2}$;
the light-matter coupling within the dipole and rotating-wave
approximations is:~\cite{haug:93,axt:98}

\begin{equation}\label{eq:hop}
  H_{\rm op} =
    -\mbox{$\frac 1 2$}{\cal E}_o \sum_{12}(
    e^{i\omega t}\;\m{21}{he}\;\d{2}\c{1}+
    e^{-i\omega t}\;\m{12}{eh}\;\cd{1}\dd{2}),
\end{equation}

\noindent with ${\cal E}_o\exp\pm i\omega t$ the electric field of the
exciting laser light and $\m{12}{eh}$ the optical dipole matrix
elements; the Coulomb terms can be split into three parts $\delta
H=\delta H^{(0)}+\delta H^{(1)}+\delta H^{(2)}$,~\cite{axt:98} where
$\delta H^{(0)}$ conserves the number of electron-hole pairs, $\delta
H^{(1)}$ changes the number of electron-hole pairs by one, and $\delta
H^{(2)}$ by two (see Appendix \ref{appendix:ham} for details).

\subsection{Density-matrix hierarchy}

Having established our model in Sec.~\ref{sec:ham}, we next discuss how
to treat Coulomb correlation effects due to $\delta H$. Within the
framework of density matrices the central quantities are the one-time
correlation functions $\langle{\cal A}\rangle_t=\mbox{tr}
({\bm\rho}_t{\cal A})$, with ${\bm\rho}_t$ the statistical operator at
time $t$, $\mbox{tr}()$ denoting the trace over a complete set of basis
functions, and ${\cal A}$ an operator consisting of $\c{}$, $\d{}$,
$\cd{}$, and $\dd{}$; for instance, $\langle\cd{}\c{}\rangle_t$
($\langle\dd{}\d{}\rangle_t$) describes the occupation of electron
(hole) single-particle states at time $t$, whereas higher-order
correlation functions, such as, e.g.,
$\langle\cd{}\dd{}\d{}\c{}\rangle_t$, account for correlations among
two or more carriers. As the central approximation within any
density-matrix description one has to restrict oneself to a limited
number of correlation functions. The temporal evolution of the
correlation functions $\langle A\rangle_t$ is then provided by
Ehrenfest's theorem:~\cite{haug:93,fick:90}

\begin{equation}\label{eq:eom}
  \partial_t\langle{\cal A}\rangle_t=\langle[{\cal A},H]\rangle_t,
\end{equation}

\noindent which, together with the restriction to a finite number of
correlation functions, provides the basis of the framework of density
matrices. Before addressing the problem of dielectric screening in
Sec.~\ref{sec:screen}, in the following we briefly review one of the
most commonly used truncation schemes in phenomenological
semiconductor optics: The {\em dynamics-controlled
truncation}.~\cite{axt:98}\/ This discussion will help us to establish
some of the basic concepts and notations.

\subsection{Dynamics-controlled truncation}\label{sec:dct}

In describing the optical properties of conventional semiconductors one
often relies on effective models. Besides a simplified bandstructure
description in terms of $\k.\p$\/-theory or effective-mass
approximations one usually:~\cite{haug:93,rossi:98} employs the
envelope-function approximation; neglects the non-pair conserving terms
$\delta H^{(1)}$ and $\delta H^{(2)}$; and screens the Coulomb
interactions in $\delta H^{(0)}$ by the static dielectric constant (see
Sec.~\ref{sec:screen} for a first-principles motivation of such
approximation). Hence, the remaining terms $\bar H_o+\delta H^{(0)}$
conserve the number of electrons and holes, and the only source for the
creation or destruction of electron-hole pairs is through the
light-matter coupling $H_{\rm op}$; more specifically, inspection of
Eq.~(\ref{eq:hop}) reveals that the creation (destruction) of
electron-hole pairs is through terms of the form ${\cal
E}_o\;\cd{}\dd{}$ (${\cal E}_o\;\d{}\c{}$). In the pioneering work of
Axt and Stahl \cite{axt:94} (see Ref.~\onlinecite{axt:98} for a review)
the authors first noted that, since any pair of field operators
$\cd{}\dd{}$ or $\d{}\c{}$ comes with an electric field ${\cal E}_o$,
the density matrices can be classified according to their power in the
electric field.

Thus, in linear optical response there is only one correlation function

\begin{equation}\label{eq:y}
  Y_{12}=\langle \d{2}\c{1} \rangle
\end{equation}

\noindent which is of lowest order in ${\cal E}_o$. $Y$ describes the
correlation between an optically excited electron and hole (note that
$Y$ is a non-equilibrium quantity which is nonzero only in presence of
an exciting laser and which vanishes in thermal equilibrium). Its
relation to the total interband polarization ${\cal P}(t)$ is given
through ${\cal P}(t)=\sum_{12}\m{21}{he}Y_{12,t}$, with $\m{}{eh}$ the
optical dipole matrix elements. Recalling that the optical absorption
is related through Maxwell's equation to the total interband
polarization ${\cal P}(t)$,~\cite{haug:93,rossi:98} one immediately
notes that the microscopic interband polarizations $Y$ completely
determine the linear optical response of semiconductors.  Their
temporal evolution is given by Eq.~(\ref{eq:eom}), and we find after
some straightforward algebra:~\cite{axt:98,rossi:98}

\begin{equation}\label{eq:eom.y}
  i\dot Y_{12}=(\e{1}{e}+\e{2}{h})Y_{12}-
  \mbox{$\frac 1 2$}{\cal E}_o e^{-i\omega t}\;\m{12}{eh}
  -(\v{1\bar 1}{2\bar 2}{ee}{hh}-\v{12}{\bar 2\bar 1}{eh}{he})\;
    Y_{\bar 1\bar 2},
\end{equation}

\noindent where we have implicitly assumed summation over barred
indices $\bar 1$ and $\bar 2$. On the right-hand side of
Eq.~(\ref{eq:eom.y}):  the first term accounts for the single-particle
states available for adding an electron-hole pair; the second term
describes the creation of electron-hole pairs through coupling to the
light field; finally, the third term accounts for the electron-hole
Coulomb interactions (with ${\rm V}^{{\rm ee},{\rm hh}}$ the direct and
${\rm V}^{{\rm eh},{\rm he}}$ the exchange term), and is responsible
for excitonic renormalizations in the optical spectra (which would be
absent in a simple-minded single-particle description).

Eq.~(\ref{eq:eom.y}) is conveniently solved by finding the polarization
eigenmodes. The homogeneous part of Eq.~(\ref{eq:eom.y}) is then
transformed to an eigenvalue problem (``excitonic eigenvalue
problem''),~\cite{haug:93,axt:98,rossi:98} whose solutions provide the
exciton energies $E_x$ and wavefunctions $\Psi_{12}^x$. In the optical
spectra the optical transition energies are then provided by $E_x$, and
the respective oscillator strengths are given by
$|\sum_{12}\Psi_{12}^x\; \m{21}{he}|^2$.


\section{Dielectric screening}\label{sec:screen}

As it is well known that dielectric screening is of central importance
for the quantitative description of the optical properties of
semiconductors, within an {\em ab-initio}\/ approach it would make
little sense to directly employ Eq.~(\ref{eq:eom.y}) with the bare
(i.e., unscreened) Coulomb matrix elements ${\rm V}$. For instance, in
conventional semiconductors, such as Si or GaAs, dielectric screening
leads to a reduction of the unscreened ${\rm V}$'s by a factor of
approximately ten. We are thus faced with the central question of this
work: How can dielectric screening be described within the framework of
density matrices?

Quite generally, dielectric screening is a process where a carrier
polarizes its surrounding medium. In turn, a second carrier not only
feels the bare Coulomb interaction exerted by the first carrier, but
also the field produced by this induced polarization cloud. In an
electron-hole picture, such polarization effects are described by
(virtual) excitations of electron-hole pairs which result in
microscopic polarization fields. Thus, a proper first-principles
description of dielectric screening requires besides the
pair-conserving Coulomb terms $\delta H^{(0)}$ also the non-pair
conserving terms $\delta H^{(1)}$ and $\delta H^{(2)}$.

Inclusion of such terms, however, spoils the concept of the
dynamics-controlled truncation of the density-matrix hierarchy, since
electron-hole pairs are no longer solely created (destroyed) by the
light field but, in addition, also by Coulomb interactions. Thus, it is
no longer possible to rigorously classify correlation functions
according to their power in the electric field ${\cal E}_o$. However,
since both the light field and the Coulomb interactions create
electrons and holes only pairwise, we can still assume that the
classification of correlation functions in numbers of electron-hole
pairs provides a meaningful concept. For simplicity, in this paper we
restrict ourselves to the case of correlation functions with at most
two electron-hole pairs, and we show that already at this level of
approximation dielectric screening is included.

\subsection{Dynamic variables}

Besides the interband polarizations $Y_{12}$, we hence need the
electron (hole) distribution function
$C_{12}=\langle\cd{2}\c{1}\rangle$ ($D_{12}=\langle\dd{2}\d{1}\rangle$)
and the two-particle correlation functions:~\cite{axt:98}

\begin{eqnarray}
  B_{12,34}&=&\langle \d{4}\c{3}\d{2}\c{1}\rangle\label{eq:b}\\
  N_{12,34}&=&\langle \cd{3}\dd{4}\d{2}\c{1}\rangle.\label{eq:n}
\end{eqnarray}

\noindent In analogy to the physical meaning of $Y$ as a measure of the
electron-hole correlation and $C$ ($D$) as the occupation and intraband
coherence of single-particle electron (hole) states, we can
approximately interpret $B$ as a measure for the coherence between two
electron-hole pairs and $N$ as the occupation of electron-hole pairs.

Next, we derive the equations of motion for the various correlation
functions. Before doing so, we employ a further approximation. From the
analysis of the nonlinear coherent optical response (i.e., the
dynamics-controlled truncation at the level of ${\cal E}_o^3$, which
involves $Y$, $C$, $D$, $B$, and $N$)~\cite{axt:98} it is known that
the main effect of the carrier distribution functions $C$ and $D$ is to
reduce the optical transition rates because of state filling (i.e.,
Pauli blocking). As we expect such effects to be of minor importance
for the problem of our present concern, in the following we shall
neglect $C$ and $D$. Thus, our set of dynamic variables is provided by
$Y$, $B$, and $N$.

\subsection{Equations of motion for $Y$, $B$, and $N$}

The program pursued in the following is the derivation of the equations
of motion for $Y$, $B$, and $N$. This is done by using Ehrenfest's
theorem, Eq.~(\ref{eq:eom}), together with the definitions for $Y$,
$B$, and $N$, Eqs.~(\ref{eq:y},\ref{eq:b},\ref{eq:n}) (see also
Appendix \ref{appendix:ham}). After some straightforward algebra we
obtain:~\cite{remark:y*}

\begin{eqnarray}\label{eq:eom.y.full}
  i\dot Y_{12} &\cong& 
    (\e{1}{e}+\e{2}{h})\;Y_{12}-
    \mbox{$\frac 1 2$}{\cal E}_oe^{-i\omega t}\m{12}{eh}\nonumber\\
    &&-
    (\v{1\bar 1}{2\bar 2}{ee}{hh}-\v{12}{\bar 2\bar 1}{eh}{he})\;
    Y_{\bar 1\bar 2}\nonumber\\
    && +(\v{1\bar 1}{34}{ee}{eh}-\v{3\bar 1}{14}{ee}{eh})\;
    N_{\bar 12,34}+\v{1\bar 1}{43}{ee}{he}\;B_{\bar 12,34}\nonumber\\
    && -(\v{2\bar 2}{34}{hh}{eh}-\v{4\bar 2}{32}{hh}{eh})\;
    N_{1\bar 2,34}-\v{2\bar 2}{43}{hh}{he}\;B_{1\bar 2,34}\quad
\end{eqnarray}

\noindent (here and henceforth we implicitly assume summation over all
single-particle indices with exception of those appearing on the
left-hand side). Comparison with Eq.~(\ref{eq:y}), which was derived by
neglecting non-pair conserving Coulomb couplings, reveals the
appearance of an additional coupling to the two-particle correlations
$B$ and $N$, which is mediated through the non-pair-conserving Coulomb
interactions $\delta H^{(1)}$ and $\delta H^{(2)}$. As will be shown in
the following, such terms are needed for a proper description of
dielectric screening.

Let us discuss this coupling between $Y$ and $B$, $N$ in slightly more
detail. Suppose that the system under investigation is a bulk
semiconductor. Hence, the single-particle states $1,2,\dots$ consist of
a wavevetor $\k$ and a band index $n$. As the light field couples
electron and hole states with approximately opposite $\k$-values, the
optically induced interband polarization is of the form $Y_\k=\langle
\d{-\k}\c{\k}\rangle$, where for notational simplicity we have dropped
all band indices. The coupling ${\rm V}B$ in Eq.~(\ref{eq:eom.y.full})
is then of the form $\sum_{\p\q}{\rm V}(\q)\langle\d{-\p+\q}\c{\p}\;
\d{-\k+\q}\c{\k}\rangle$, with a corresponding expression for ${\rm
V}N$. Hence, these terms describe how the propagation of $Y$ is
modified by the presence of polarization fluctuations in the system; as
we will show in the following, such fluctuations are induced by the
interband polarization $Y$, thus resulting in a self-interaction-like
process where: the interband polarization $Y$ induces polarization
fluctuations through the non-pair-conserving Coulomb couplings $\delta
H^{(1)}$; these fluctuations propagate in time, which is described by
the equations of motion for $B$ and $N$ (to be derived below); and
finally couple back to $Y$, which is described by the terms ${\rm V}B$
and ${\rm V}N$ in Eq.~(\ref{eq:eom.y.full}).

Next, we derive the equations of motion for $B$ and $N$. To keep our
analysis as simple as possible, we employ from the beginning a number
of well-controlled approximations. First, we assume that before arrival
of the exciting laser no two-particle correlations $B$ and $N$ are
present (as will be discussed below this is only an approximation).
Hence, $Y$, $B$, $N$ are induced by the light-semiconductor coupling
$H_{\rm op}$, and in linear response it suffices to keep in the
dynamics only terms linear in $Y$, $B$, $N$ (i.e., we neglect
contributions proportional to, e.g., $Y^2$ or $YB$). Second, we neglect
the light coupling in the equations of motion for $B$ and $N$ (such
terms would describe interference terms between dielectric screening
and light coupling). Third, we shall employ the so-called random-phase
approximation (RPA):~\cite{negele:88,hohenester:97} For $B_{12,34}$ and
$N_{12,34}$ we keep only Coulomb terms which scatter one pair of
particles $12$ ($34$) and leave the other pair $34$ ($12$) unaffected.
In other words, we assume that polarization fluctuations created
through light-coupling and Coulomb processes, respectively, move
independently of each other. In a $k$-space representation for the
polarization fluctuations
$\langle\d{-\p+\q}\c{\p}\;\d{-\k+\q}\c{\k}\rangle$ and
$\langle\cd{\p}\dd{-\p+\q}\;\d{-\k+\q}\c{\k}\rangle$ one readily
observes that these RPA terms contain all those contributions which
involve Coulomb matrix elements ${\rm V}(\q)$ with a momentum exchange
of $\q$ which is independent of $\k$ and $\p$.~\cite{hohenester:97}

With these approximations we then find:

\begin{widetext}
\begin{eqnarray}
  i\dot B_{12,34}&\cong&
    (\e{1}{e}+\e{2}{h}+\e{3}{e}+\e{4}{h})\;B_{12,34}\nonumber\\
    &+&(\v{12}{\bar 2\bar 1}{eh}{he}-\v{1\bar 1}{2\bar 2}{ee}{hh})\;
    B_{\bar 1\bar 2,34}+
    (\v{12}{\bar 1\bar 2}{eh}{eh}-\v{1\bar 2}{\bar 12}{eh}{eh})\;
    N_{34,\bar 1\bar 2}\nonumber\\
    &+&(\v{34}{\bar 4\bar 3}{eh}{he}-\v{3\bar 3}{4\bar 4}{ee}{hh})\;
    B_{12,\bar 3\bar 4}+
    (\v{34}{\bar 3\bar 4}{eh}{eh}-\v{3\bar 4}{\bar 34}{eh}{eh})\;
    N_{12,\bar 3\bar 4}\nonumber\\ 
    &+&(\v{1\bar 1}{34}{ee}{eh}-\v{3\bar 1}{14}{ee}{eh})\;Y_{\bar 12}-
    (\v{2\bar 2}{34}{hh}{eh}-\v{4\bar 2}{32}{hh}{eh})\;Y_{1\bar 2}
    \nonumber\\
    &+&(\v{3\bar 3}{12}{ee}{eh}-\v{1\bar 3}{32}{ee}{eh})\;Y_{\bar 34}-
    (\v{4\bar 4}{12}{hh}{eh}-\v{2\bar 4}{14}{hh}{eh})\;Y_{3\bar 4}+
    (\v{12}{34}{eh}{eh}-\v{14}{32}{eh}{eh})\label{eq:eom.b}\\
    i\dot N_{12,34}&\cong&
    (\e{1}{e}+\e{2}{h}-\e{3}{e}-\e{4}{h})\;N_{12,34}\nonumber\\
    &+&(\v{12}{\bar 2\bar 1}{eh}{he}-\v{1\bar 1}{2\bar 2}{ee}{hh})\;
    N_{\bar 1\bar 2,34}+
    (\v{12}{\bar 1\bar 2}{eh}{eh}-\v{1\bar 2}{\bar 12}{eh}{eh})\;
    B_{\bar 1\bar 2,34}^*\nonumber\\
    &-&(\v{43}{\bar 3\bar 4}{he}{eh}-\v{\bar 33}{\bar 44}{ee}{hh})\;
    N_{12,\bar 3\bar 4}-
    (\v{43}{\bar 4\bar 3}{he}{he}-\v{\bar 43}{4\bar 3}{he}{he})\;
    B_{12,\bar 3\bar 4}.\label{eq:eom.n}
\end{eqnarray}
\end{widetext}

\noindent Let us analyze the various contributions to these equations
in slightly more detail. On the right-hand side of
Eq.~(\ref{eq:eom.b}): the first term corresponds to the free
propagation of $B$; the terms in the second and third line,
respectively, describe Coulomb renormalization processes in the
propagation of $B$ (${\rm V}B$) and Coulomb couplings between $N$ and
$B$ (${\rm V}N$); the terms in the last two lines are the source terms
which describe how polarization fluctuations are created through
coupling to the interband polarizations $Y$. In addition, there is a
term $(\v{12}{34}{eh}{eh}-\v{14}{32}{eh}{eh})$ which describes the
buildup of polarization fluctuations even in absence of light
couplings. Quite generally, its appearance is not unexpected since up
to now we have assumed that the semiconductor groundstate is obtained
by simply filling DFT single-particle states. Acting with $\delta
H^{(2)}$ on the DFT vacuum, however, we immediately observe that Coulomb
interactions create electron-hole pairs and thus lead to
renormalizations of the DFT vacuum. As a first approximation, in this
paper we assume that such renormalizations are not of crucial
importance for the description of dynamic processes in the propagation
of $Y$, and we thus neglect this term. 

In the equation of motion for $N$, Eq.~(\ref{eq:eom.n}), we observe in
analogy to the dynamics of $B$ single-particle contributions (first
line) and Coulomb renormalizations and couplings (second and third
line), but no source terms are present. Similarly to the semiconductor
Bloch equations,~\cite{haug:93,rossi:98} which describe the
non-equilibrium optical response at the level of two-point functions
(i.e., $C$, $D$, and $Y$) and where the light field drives the
electron-hole coherence $Y$ which, in turn, acts as a source term for
the electron (hole) occupations $C$ ($D$), we here have the situation
that Coulomb coupling first drives the electron-hole pair coherence $B$
which, in turn, acts as source for the electron-hole pair occupation
$N$.

Eqs.~(\ref{eq:eom.y.full}--\ref{eq:eom.n}) are the central equations of
this work. Before showing how to solve this set of equations, we
introduce two further approximations. First, we keep in
Eqs.~(\ref{eq:eom.b},\ref{eq:eom.n}) only self-interaction like
processes. The electron-hole coherence $Y_{\bar 12}$ ($Y_{1\bar 2}$)
between the states $\bar 12$ ($1\bar 2$) is initially created through
the coupling to the light field. In a self-interaction process: this
coherence is transferred through Coulomb coupling of the electron
(hole) to $12$ and to a second electron-hole pair $34$ [fourth line in
Eq.~(\ref{eq:eom.b})]; the second electron-hole pair $34$ propagates in
presence of Coulomb renormalizations [third lines in
Eqs.~(\ref{eq:eom.b},\ref{eq:eom.n})] and the initial pair $12$ remains
unscattered; finally, this coherence between electron-hole pairs $12$
and $34$ affects the propagation of $Y$ [third and fourth line in
Eq.~(\ref{eq:eom.y.full})]. In addition to these self-interaction like
processes, there exist also scattering-like contributions: here either
the optically induced coherence between electron-hole pair $34$ is
transferred through Coulomb coupling to the pair $12$ [fifth line in
Eq.~(\ref{eq:eom.b})] or the coherence between pair $12$ is scattered
to $\bar 1\bar 2$ [second lines in
Eqs.~(\ref{eq:eom.b},\ref{eq:eom.n})]. Such terms, which will be
neglected in the following, describe higher-order Coulomb terms (within
the framework of Green's functions they would correspond to vertex
corrections); a more detailed discussion of such scattering
contributions has been given in Ref.~\onlinecite{hohenester:97}.

Our second approximation concerns the neglect of electron-hole exchange
interactions in the dynamics of $B$ and $N$: in
Eqs.~(\ref{eq:eom.b},\ref{eq:eom.n}) we only keep the first Coulomb
terms in parentheses. One readily observes in a $k$-space representation
that for the polarization fluctuations $\langle\d{-\p+\q}\c{\p}\;
\d{-\k+\q}\c{\k}\rangle$ and $\langle\cd{\p}\dd{-\p+\q}\;
\d{-\k+\q}\c{\k}\rangle$ these terms correspond to Coulomb interactions
${\rm V}(\q)$ where the exchanged momentum $\q$ is {\em independent}\/
of the momenta $\k$ and $\p$; the remaining terms involve exchanged
momenta depending on $\k$ and $\p$. The more general expressions of
Eqs.~(\ref{eq:eom.b},\ref{eq:eom.n}) which contain both the direct and
exchange contributions have been given for the purpose of our later
discussion about the screening of the electron-hole exchange
interaction.

Details of our solution scheme for Eqs.~(\ref{eq:eom.b}) and
(\ref{eq:eom.n}) are presented in Appendix
\ref{appendix:details.solution}. The key to the solution is the
Dyson-like character of the equations of motion for $B$ and $N$. In
fact, solving Eqs.~(\ref{eq:eom.b},\ref{eq:eom.n}) by iteration we
observe that the repeated action of ${\rm V}(B+N)$ precisely
reproduces Dyson's equation for the screened Coulomb potential ${\rm
W}={\rm V}[1+{\rm V}P+({\rm V}P)^2+\dots]={\rm V}+{\rm W}P{\rm V}$;
here $P$ is the usual retarded polarization function within
random-phase approximation. Finally, we arrive at (for details see
Appendix \ref{appendix:details.solution}):

\begin{widetext}
\begin{eqnarray}\label{eq:eom.y.final}
  i\dot Y_{12} &\cong&
    (\e{1}{e}+\e{2}{h})\;Y_{12}+
    \D{1\bar 1}{\bar 1\tilde 1}{ee}{ee}(\omega_{\bar 12})
    \;Y_{\tilde 12}+
    \D{2\bar 2}{\bar 2\tilde 2}{hh}{hh}(\omega_{1\bar 2})
    \;Y_{1\tilde 2}-
    \mbox{$\frac 1 2$}{\cal E}_oe^{-i\omega t}\m{12}{eh}\nonumber\\
    &&-\bigl(
    \v{1\bar 1}{2\bar 2}{ee}{hh}+
    \D{1\bar 1}{2\bar 2}{ee}{hh}(\omega_{\bar 12})+
    \D{2\bar 2}{1\bar 1}{hh}{ee}(\omega_{1\bar 2})-
    \v{12}{\bar 2\bar 1}{eh}{he}\bigr)\;
    Y_{\bar 1\bar 2}\nonumber\\
\end{eqnarray}
\end{widetext}

\noindent with

\begin{equation}\label{eq:d}
  {\cal D}(\r,\r';\omega)=\int d(\s\s'\bar\omega)\;
  {\rm W}(\r,\s;\omega)\frac{D_o(\s,\s';\bar\omega)}%
                            {\omega-\bar\omega+i0^+}
  {\rm V}(\s',\r'),
\end{equation}

\noindent [the relation between the real-space and single-particle
representations of ${\cal D}$ is in analogy to
Eq.~(\ref{eq:def.coul})]; $D_o$ is related to the imaginary part of the
polarization function through $D_o(\r,\r';\omega)=-\pi^{-1}\Im
P(\r,\r';\omega>0)$ [Eqs.~(\ref{eq:pol.a},\ref{eq:pol.b})]. A
particularly simple expression of Eq.~(\ref{eq:eom.y.final}) follows
for static screening, i.e., for:

\begin{equation}\label{eq:d.static}
  {\cal D}(\r,\r';\omega=0)=\mbox{$\frac 1 2$} 
  [{\rm W}(\r,\r';0)-{\rm V}(\r,\r')], 
\end{equation}

\noindent where the expression in the second line of
Eq.~(\ref{eq:eom.y.final}) reduces to $[\w{1\bar 1}{2\bar
2}{ee}{hh}(0)- \v{12}{\bar 2\bar 1}{eh}{he}]\;Y_{\bar 1\bar 2}$. Thus,
in Eq.~(\ref{eq:eom.y.final}) the ${\cal D}$-terms in the first line
describe renormalization effects in the propagation of single electrons
and holes, respectively, and the terms in the second line account for
the Coulomb coupling between the optically excited electron-hole pair.


\section{Discussion}

Equation (\ref{eq:eom.y.final}) is our final expression. Before
analyzing it in more detail, let us briefly recall the approximations
that were adopted in its derivation. First, within the spirit of the
dynamics-controlled truncation~\cite{axt:98} only dynamic variables
accounting at most for two electron-hole pairs were considered. The
electron and hole distribution functions $C$ and $D$ were neglected,
since from related work it is known that their main influence is the
blocking of optical transitions due to state filling. Thus, our set of
dynamic variables consists of the interband polarizations $Y$, and the
two-particle correlations $B$ and $N$ which account for the coherence
between electron-hole pairs and their occupation, respectively. We
assumed that before arrival of the exciting laser light $Y$, $B$, and
$N$ vanish. In the equations of motion for $B$ and $N$: we adopted the
random-phase approximation~\cite{hohenester:97} which assumes that
polarization fluctuations propagate independently of each other; we
neglected the direct coupling of $B$ and $N$ to the light field; only
self-interaction-like processes were kept; finally, exchange-type
interactions were neglected.

Quite generally, none of these approximations is compulsory, and all of
them could be lifted without introducing major conceptual
modifications. However, we have kept our analysis as simple as possible
in order to emphasize the two primary goals of this work: first, the
identification of those terms and approximations which are required
within a density-matrix formalism in order to describe dielectric
screening; and second, to discuss the respective differences between
our approach and the combined GWA and Bethe-Salpeter equation
approach.

Within our density-matrix description, the central quantities are the
optically induced interband polarizations $Y$. In general, their
knowledge allows the full calculation of the optical absorption
spectra. However, as discussed in Sec.~\ref{sec:dct}, the temporal
evolution of $Y$ is in addition governed by the two-particle
correlations $B$ and $N$, which, within a self-interaction like
process, are induced through the non-pair conserving Coulomb couplings
$\delta H^{(1)}$ and $\delta H^{(2)}$; the details of this coupling,
Eqs.~(\ref{eq:eom.y.full}--\ref{eq:eom.n}), reflect the essential
features of dielectric screening where: an electron (hole) polarizes
its surrounding medium (described by the source terms for $B$); in
turn, the hole (electron) not only feels the bare Coulomb term exerted
by the the first carrier but also the field induced by the polarization
cloud [cf. terms in the second line of Eq.~(\ref{eq:eom.y.final})]; in
addition, this induced polarization cloud also affects the propagation
of the carrier [cf. ${\cal D}$-terms in the first line of
Eq.~(\ref{eq:eom.y.final})]. Thus, the screening of the electron-hole
interaction and the quasiparticle renormalizations originate from the
{\em same}\/ physical process.

It is interesting to compare this finding with the respective GWA and
Bethe-Salpeter equation result. Let us first concentrate on the case of
static screening, Eq.~(\ref{eq:d.static}). Inspection of the
quasiparticle renormalizations of Eq.~(\ref{eq:eom.y.final}) with the
results of the GWA approach~\cite{aryasetiawan:98} reveals that these
renormalizations closely resemble the screened exchange and Coulomb
hole self-energy terms (note that this comparison is somewhat
complicated because of our use of the electron-hole picture and the
missing screening of the electron-hole exchange interactions). If
dynamic screening is considered, the inelasticities $\omega_{12}$ of
${\cal D}(\omega_{12})$ in Eq.~(\ref{eq:eom.y.final}) are given by the
differences between the light frequency $\omega$ and the electron-hole
transition energies $\e{1}{e}+\e{2}{h}$. Assuming that the exciton
states are composed of electron and hole single-particle states with
energies close to the bandgap, one immediately notes that
$\omega\approx\e{1}{e}+\e{2}{h}$; thus, static screening of the direct
Coulomb interactions is expected to be a good approximation. This
result is different with respect to the GWA prescription, where the
inelasticities of the quasiparticle renormalizations [first line of
Eq.~(\ref{eq:eom.y.final})] are the quasiparticle energies themselves,
and the electron-hole interaction is screened at the optical
frequency~\cite{rohlfing:00}, $\omega$ (see also
Ref.~\onlinecite{bechstedt:97} for a discussion of the compensation
between effects due to dynamical screening and higher-order vertex
corrections; the consequences of such interplay within our present
density-matrix description will be discussed elsewhere).

Next, we discuss the screening of the electron-hole exchange
interactions ${\rm V}^{\rm eh,he}$. As we saw in the discussion of the
screening of the direct Coulomb terms, dielectric screening originates
from a process where: the interband polarizations $Y$ induce the
electron-hole pair correlations $B$ and $N$; these induced polarization
fluctuations move independently of each other (random-phase
approximation); and finally couple back to the propagation dynamics of
$Y$. These self-interaction-like processes result in the terms ${\cal
D}_{1\bar 1,2\bar 2}$, Eq.~(\ref{eq:d}), where the vertices $2\bar 2$
and $1\bar 1$, respectively, reflect the coupling of $Y$ to the
electron-hole pair coherence $B$ and the back action of $B$, $N$ on
$Y$; finally, the propagation dynamics of $B$ and $N$ is hidden in
${\cal D}$, which has to be determined from the Dyson-like equation for
the screened Coulomb potential ${\rm W}$. Let us now return to the full
expression of Eqs.~(\ref{eq:eom.b},\ref{eq:eom.n}) which, within the
approximations adopted, contain all possible source terms. One
immediately recognizes that neither these source terms nor the coupling
of $B$, $N$ to $Y$, Eq.~(\ref{eq:eom.y.full}), can reproduce a
screening term of the form ${\cal D}^{\rm eh,he}$. Hence, within our
approximation scheme (RPA) the electron-hole exchange interactions must
remain unscreened. Future work will address possible screening
contributions and vertex corrections beyond the random-phase
approximation.

Finally, we briefly comment on one of the well-known shortcomings of
our density-matrix description. From Eqs.~(\ref{eq:eom.y.final}) and
(\ref{eq:pol.a}) one readily observes that the {\em unrenormalized}\/
single-particle energies $\e{}{e}$ and $\e{}{h}$ enter into the
calculation of the screened Coulomb potential and of the polarization
function $P$. This result differs from the corresponding GWA and
Bethe-Salpeter equation result and is due to our neglect of
higher-order correlation functions (i.e., correlations between three or
more electron-hole pairs). However, it can be shown~\cite{bonitz:98}
that inclusion of certain types of Coulomb interactions at this level
of many-particle correlations indeed results in a renormalization of
the single-particle energies $\e{}{e}$ and $\e{}{h}$. For the sake of
brevity, here we shall not present the details of such analysis.

When defining the Coulomb interactions of our starting Hamiltonian,
Eq.~(\ref{eq:ham.coul(psi)}), we assumed that exchange interactions are
fully accounted for within the DFT calculations. Apparently, this
assumption no longer holds for DFT calculations based on the celebrated
local-density approximation (LDA).~\cite{dreizler:90}. Inspection of
Eq.~(\ref{eq:ham.coul(psi)}) reveals that in this case the substracted
exchange contributions would be of the form $v_x(\r,\r')\cong
\delta(\r-\r')v_x^{\mbox{\tiny LDA}}(\r)$; as consequence, in $\delta
H$ only the Hartree contributions with the filled valence band would be
cancelled exactly (with no corresponding cancellation for the exchange
terms) and the local exchange potential would give rise to terms
proportional to $\cd{}\dd{}$ and $\d{}\c{}$. In the equations of motion
for the dynamic variables these local-exchange contributions would give
rise to additional couplings between variables containing different
numbers of electron-hole pairs. Within the spirit of perturbation
theory, such couplings are due to the fact that the semiconductor
groundstate is not given by simply filling the DFT--LDA states; rather,
the non-local exchange interactions give rise to admixtures of excited
DFT--LDA states, which, within the electron-hole picture, correspond to
dynamic variables accounting for different numbers of electron-hole
pairs. However, it is well known that in many cases the single-particle
renormalizations to the DFT--LDA states simply lead to a rigid shift of
the calculated bandgap (scissor operator) and that the calculated
wavefunctions remain unchanged. In this case, the results of our
analysis remain valid and only the electron-hole exchange interactions
${\rm V}^{\rm eh,he}$, which correct for the missing Coulomb exchange
interactions because of optical excitation, should be computed within
the local-density approximation.

\section{Conclusions and Summary}

In conclusion, starting from an {\em ab-initio}\/ bandstructure
calculation we have shown how to describe dielectric screening within a
density-matrix description. We have identified those dynamic variables
and approximations which are required at the lowest level of
approximation (random-phase approximation). We have discussed that
static screening is expected to be a good approximation (in contrast to
the corresponding Green's function results~\cite{rohlfing:00} which
seem to favour screening at the optical frequencies). Hence, our
analysis suggests calculation of the optical absorption spectra from
the solutions of the ``excitonic eigenvalue
problem'':~\cite{axt:98,haug:93,rossi:98}

\begin{equation}\label{eq:prescription}
  (\bar\epsilon_1^{\;\rm e}+\bar\epsilon_2^{\;\rm h})\Psi_{12}^x-
  (\w{1\bar 1}{2\bar 2}{ee}{hh}(0)-\v{12}{\bar 2\bar 1}{eh}{he})
  \Psi_{\bar 1\bar 2}^x=E_x\Psi_{12}^x,
\end{equation}

\noindent with $E_x$ and $\Psi^x$ the exciton energies and
wavefunctions, respectively (with proper summation over barred
indices).  Here, we have assumed that single-particle renormalizations
(scissor operator, GW corrections) lead to a rigid shift of the
calculated bandgap, with $\bar\epsilon_1^{\;\rm e,h}$ the renormalized
single-particle energies [see Eq.~(\ref{eq:eom.y.final}) for the more
complete expression]; the ${\rm W}$'s are the Coulomb interactions
screened within RPA, where the dielectric function is computed at zero
frequency (static-screening approximation) using the renormalized
single-particle energies $\bar\epsilon_1^{\;\rm e,h}$; finally, ${\rm
V}^{\rm eh,he}$ are the unscreened electron-hole exchange interactions;
within LDA, these terms should be computed within the local
approximation. In the optical spectra, the optical transition energies
are provided by $E_x$, and the respective oscillator strengths are
given by $|\sum_{12}\Psi_{12}^x\; \m{21}{he}|^2$.

It is hoped that our approach might serve as an alternative to the
hitherto used GWA and Bethe-Salpeter-equation approaches. Its advantage
lies in its conceptual simplicity: only some basic knowledge about
statistical mechanics and Heisenberg's equations of motion are required
(as opposed to the more sophisticated framework of Green's functions).
Finally, this work might help to extend the variety of techniques
developed recently within the field of semiconductor optics for the
description of the nonlinear optical response to first-principles
studies.

\acknowledgements

I gratefully acknowledge most helpful discussions with Alice Ruini,
Marilia Caldas, and Elisa Molinari.


\begin{appendix}

\section{}\label{appendix:ham}

In this Appendix we discuss the transformation of
Eq.~(\ref{eq:ham.coul(psi)}) to the electron-hole picture. The relation
between the field operators ${\bm\psi}$ and $\c{}$, $\d{}$ is given
by:

\begin{equation}\label{eq:field.op}
  {\bm\psi}^\dagger(\r)=\sum_1 [\phi_1^{\rm e}(\r)]^*\cd{1}+
                        \sum_2 \phi_2^{\rm h}(\r)\d{2}
\end{equation}	

\noindent Inserting this expression into Eq.~(\ref{eq:ham.coul(psi)}),
we obtain after some straightforward algebra:

\begin{widetext}
\begin{subequations}\label{eq:ham.coul}
\begin{eqnarray}
  \delta H^{(0)} &=&
    \mbox{$\frac 1 2$}(
    \v{\bar 11}{\bar 22}{ee}{ee}\;\cd{\bar 1}\cd{\bar 2}\c{1}\c{2}+
    \v{\bar 11}{\bar 22}{hh}{hh}\;\dd{\bar 1}\dd{\bar 2}\d{1}\d{2})-
    (\v{\bar 11}{\bar 22}{ee}{hh}-\v{\bar 1\bar 2}{21}{eh}{he})\;
    \cd{\bar 1}\dd{\bar 2}\d{2}\c{1}\\
  \delta H^{(1)} &=&
    \v{12}{\bar 33}{eh}{ee}\;\cd{1}\dd{2}\cd{\bar 3}\c{3}-
    \v{12}{\bar 33}{eh}{hh}\;\cd{1}\dd{2}\dd{\bar 3}\d{3}+
    \v{\bar 33}{21}{ee}{he}\;\cd{\bar 3}\c{3}\d{2}\c{1}-
    \v{\bar 33}{21}{hh}{he}\;\dd{\bar 3}\d{3}\d{2}\c{1}\\
  \delta H^{(2)} &=&
  \mbox{$\frac 1 2$}(
  \v{12}{34}{eh}{eh}\;\cd{1}\cd{3}\dd{4}\dd{2}+
  \v{21}{43}{he}{he}\;\d{2}\d{4}\c{3}\c{1})
\end{eqnarray}
\end{subequations} 

\noindent (we assume an implicit summation over all single-particle
indices); the commutation relations between $\c{}$, $\d{}$ and the
various Coulomb terms are:

\begin{subequations}
\begin{eqnarray}
  {[}\c{1},\delta H^{(0)}{]} &=&
    \v{1\bar 1}{\bar 33}{ee}{ee}\;\cd{\bar 3}\c{3}\c{\bar 1}-
    (\v{1\bar 1}{\bar 33}{ee}{hh}-\v{1\bar 3}{3\bar 1}{eh}{he})\;
    \dd{\bar 3}\d{3}\c{\bar 1}\\
  {[}\d{2},\delta H^{(0)}{]} &=&
     \v{2\bar 2}{\bar 33}{hh}{hh}\;\dd{\bar 3}\d{3}\d{\bar 2}-
    (\v{2\bar 2}{\bar 33}{hh}{ee}\;-\v{2\bar 3}{3\bar 2}{he}{eh})\;
    \cd{\bar 3}\c{3}\d{\bar 2}\\ 
  {[}\c{1},\delta H^{(1)}{]} &=&
    \phantom{-}(\v{1\bar 1}{34}{ee}{eh}-\v{3\bar 1}{14}{ee}{eh})\;
    \cd{3}\dd{4}\c{\bar 1}-
    \v{1\bar 2}{\bar 33}{eh}{hh}\;\dd{\bar 2}\dd{\bar 3}\d{3}+
    \v{1\bar 1}{43}{ee}{he}\;\d{4}\c{3}\c{\bar 1}\\
  {[}\d{2},\delta H^{(1)}{]} &=&
    -(\v{2\bar 2}{34}{hh}{eh}-\v{4\bar 2}{32}{hh}{eh})\;
    \cd{3}\dd{4}\d{\bar 2}-
    \v{\bar 12}{\bar 33}{eh}{ee}\;\cd{\bar 1}\cd{\bar 3}\c{3}+
    \v{2\bar 2}{43}{hh}{he}\;\d{4}\c{3}\d{\bar 2}\\
  {[}\c{1},\delta H^{(2)}{]} &=& \phantom{-}
    \v{1\bar 2}{34}{eh}{eh}\;\cd{3}\dd{4}\dd{\bar 2}\\
  {[}\d{2},\delta H^{(2)}{]} &=& -
    \v{\bar 12}{34}{eh}{eh}\;\cd{3}\dd{4}\cd{\bar 1}
\end{eqnarray}
\end{subequations}
\end{widetext}

\noindent Finally, the Coulomb matrix elements in
Eq.~(\ref{eq:ham.coul}) are defined as:

\begin{equation}\label{eq:def.coul}
  \v{\bar 11}{\bar 22}{\bar \mu\mu}{\bar \nu\nu}=
  \int d(\r\r')\;\frac 1 {|\r-\r'|}
  \Phi_{\bar 11}^{\rm\bar \mu\mu}(\r)
  \Phi_{\bar 22}^{\rm\bar \nu\nu}(\r'),
\end{equation}

\noindent with

\begin{subequations}
\begin{eqnarray}
  \Phi_{\bar 11}^{\rm ee}(\r)&=&
  [\phi_{\bar 1}^{\rm e}(\r)]^*\phi_1^{\rm e}(\r)\\ 
  \Phi_{\bar 11}^{\rm hh}(\r)&=&
  [\phi_{\bar 1}^{\rm h}(\r)]^*\phi_1^{\rm h}(\r)\\
  \Phi_{12}^{\rm eh}(\r)&=&
  [\phi_1^{\rm e}(\r)\phi_2^{\rm h}(\r)]^*\\
  \Phi_{21}^{\rm he}(\r)&=&
  \phantom{[}\phi_2^{\rm h}(\r)\phi_1^{\rm e}(\r).
\end{eqnarray}
\end{subequations}

\section{}\label{appendix:details.solution}

In this Appendix we sketch the derivation of our final expression
(\ref{eq:eom.y.final}). Without loss of generality, in linear response
we consider a monofrequent laser excitation of the form ${\cal
E}_oe^{-i\omega t}$. Since $Y$, $B$, and $N$ are driven by the light
field, their time dependence is also $e^{-i\omega t}$. Hence, keeping
only self-interaction like processes and direct Coulomb terms, we find
for $B$ and $N$ the Dyson-like equations:

\begin{eqnarray}\label{eq:eom.z}
  B_{12,34}&=&\phantom{-}\frac%
  {\v{34}{\bar 4\bar 3}{eh}{he}\;B_{12,\bar 3\bar 4}+
   \v{34}{\bar 3\bar 4}{eh}{eh}\;N_{12,\bar 3\bar 4}}
  {\omega_{12}-\e{3}{e}-\e{4}{h}+i0^+}
  +B_{12,34}^0\nonumber\\ 
  N_{12,34}&=&-\frac%
  {\v{43}{\bar 4\bar 3}{he}{he}\;B_{12,\bar 3\bar 4}+
   \v{43}{\bar 3\bar 4}{he}{eh}\;N_{12,\bar 3\bar 4}}
  {\omega_{12}+\e{3}{e}+\e{4}{h}+i0^+},
\end{eqnarray}

\noindent with $\omega_{12}=\omega-\e{1}{e}-\e{2}{h}$ and
\begin{widetext}

\begin{equation}
  B_{12,34}^0=\frac%
  {\v{1\bar 1}{34}{ee}{eh}\;Y_{\bar 12}-
   \v{2\bar 2}{34}{hh}{eh}\;Y_{1\bar 2}}
   {\omega_{12}-\e{3}{e}-\e{4}{h}+ i0^+}.
\end{equation}

\noindent Here, the infinitesimal imaginary part $i0^+$ has been
introduced to ensure causality. Next, it turns out to be convenient to
introduce the mixed representations $B_{12}(\r)=\sum_{34}\Phi_{34}^{\rm
eh}(\r)\;B_{12,34}$ and $N_{12}(\r)=\sum_{34}\Phi_{43}^{\rm
he}(\r)\;N_{12,34}$. The coupling between $B$, $N$ and $Y$ [third and
fourth line in Eq.~(\ref{eq:eom.y.full})] can then be cast to the
form:

\begin{equation}\label{eq:coupling.y.bn}
  \int d(\r\r')\;\frac 1 {|\r-\r'|}\bigl(
  \Phi_{1\bar 1}^{\rm ee}(\r)[B_{\bar 12}(\r')+N_{\bar 12}(\r')]-
  \Phi_{2\bar 2}^{\rm hh}(\r)[B_{1\bar 2}(\r')+N_{1\bar 2}(\r')]
  \bigr)
\end{equation}

\noindent (note our neglect of exchange interactions). The sum of $B$
and $N$, which appears on the right-hand side of
Eq.~(\ref{eq:coupling.y.bn}), is obtained by summing
Eqs.~(\ref{eq:eom.z}) and transforming to the mixed representation for
$B$ and $N$. We obtain:

\begin{equation}\label{eq:b+n}
  B_{12}(\r)+N_{12}(\r)=\int d(\s\s')\; P(\r,\s;\omega_{12})
  \frac 1 {|\s-\s'|}[B_{12}(\s')+N_{12}(\s')]+
  B_{12}^0(\r),
\end{equation}

\noindent with the polarization in random-phase approximation:

\begin{equation}\label{eq:pol.a}
  P(\r,\r';\omega)=\sum_{34}\biggl[
  \frac{\Phi_{34}^{\rm eh}(\r)\Phi_{43}^{\rm he}(\r')}
       {\omega-\e{3}{e}-\e{4}{h}+i0^+}-
  \frac{\Phi_{43}^{\rm he}(\r)\Phi_{34}^{\rm eh}(\r')}
       {\omega+\e{3}{e}+\e{4}{h}+i0^+}
  \biggr].
\end{equation}

One readily recognizes that Eq.~(\ref{eq:b+n}) closely resembles
Dyson's equation for the screened Coulomb potential ${\rm W}$, which,
in shorthand notation, reads ${\rm W}={\rm V}+{\rm W}P{\rm V}$.
Together with:

\begin{equation}
  {\rm W}(\r,\r';\omega)=\int d\s\;\frac 1 {|\r-\s|}K(\s,\r';\omega),
\end{equation}

\noindent where $K$ is the inverse dielectric
function,~\cite{hohenester:97} we immediately obtain $(B+N)=KB^0$.
Inserting this expression into Eq.~(\ref{eq:coupling.y.bn}) we find
after some straightforward calculations:

\begin{eqnarray}
  i\dot Y_{12} &\cong& 
    (\e{1}{e}+\e{2}{h})\;Y_{12}-
    \mbox{$\frac 1 2$}{\cal E}_oe^{-i\omega t}\m{12}{eh}-
    (\v{1\bar 1}{2\bar 2}{ee}{hh}-\v{12}{\bar 2\bar 1}{eh}{he})\;
    Y_{\bar 1\bar 2}\nonumber\\
    &&+\w{1\bar 1}{34}{ee}{eh}(\omega_{\bar 12})
    \frac 1 {\omega_{\bar 12}-\e{3}{e}-\e{4}{h}+i0^+}
    \v{34}{\bar 1\tilde 1}{eh}{ee}\;Y_{\tilde 12}+
    \w{2\bar 2}{34}{hh}{eh}(\omega_{1\bar 2})
    \frac 1 {\omega_{1\bar 2}-\e{3}{e}-\e{4}{h}+i0^+}
    \v{34}{\bar 2\tilde 2}{eh}{hh}\;Y_{1\tilde 2}\nonumber\\
    &&-
    \w{1\bar 1}{34}{ee}{eh}(\omega_{\bar 12})
    \frac 1 {\omega_{\bar 12}-\e{3}{e}-\e{4}{h}+i0^+}
    \v{34}{2\bar 2}{eh}{hh}\;Y_{\bar 1\bar 2}-
    \w{2\bar 2}{34}{hh}{eh}(\omega_{1\bar 2})
    \frac 1 {\omega_{1\bar 2}-\e{3}{e}-\e{4}{h}+i0^+}
    \v{34}{1\bar 1}{eh}{ee}\;Y_{\bar 1\bar 2},\nonumber\\
\end{eqnarray}

\noindent with the matrix elements of ${\rm W}$ defined in analogy to
Eq.~(\ref{eq:def.coul}). Finally, in case of time-reversal symmetry the
wavefunctions can be chosen real. Hence, the polarization $P$ is of the
form:~\cite{aryasetiawan:98}

\begin{equation}\label{eq:pol.b}
  P(\r,\r';\omega)=\int_0^\infty d\omega'\;D_o(\r,\r';\omega')
  \bigl[\frac 1{\omega-\omega'+i0^+}-
        \frac 1{\omega+\omega'+i0^+}\bigr],
\end{equation}

\end{widetext}

\noindent with $D_o(\r,\r';\omega)=\sum_{34}\Phi_{34}^{\rm
eh}(\r)\Phi_{43}^{\rm he}(\r')\delta(\omega-\e{3}{e}-\e{4}{h})$, and we
arrive after some straightforward algebra at Eq.~(\ref{eq:eom.y.final}).

\end{appendix}


\begin{thebibliography}{10}
\bibitem{remark:y*}
\bibinfo{note}{In the spirit of the rotating-wave approximations, we neglect in
  the dynamics of $Y$ counter-clockwise rotating contributions proportional to
  $Y^*$.}
\end{thebibliography}
\end{document}